\documentclass[12pt,preprint,notoc]{QCEMDA}  

\usepackage{epsfig,multicol,bbm}

\usepackage{verbatim}

\newcommand\fverb{\setbox\pippobox=\hbox\bgroup\verb}
\newcommand\fverbdo{\egroup\medskip\noindent%
            \fbox{\unhbox\pippobox}\ }
\newcommand\fverbit{\egroup\item[\fbox{\unhbox\pippobox}]}
\newbox\pippobox

\title{Thermodynamics of accelerating and rotating black holes}

\author{Muhammad Bilal and K. Saifullah\\

Department of Mathematics, Quaid-i-Azam University, Islamabad,
Pakistan \\

Electronic address: \email{saifullah@qau.edu.pk}}

\preprint{}  

\abstract{Thermodynamics of a large family of black holes from
electrovacuum solutions of Einstein's equations is studied. This
family includes rotating and non-accelerating black holes with NUT
charge, and rotating and accelerating black holes. The surface
gravity, Hawking temperature and the area laws for these black holes
are presented. The first law of thermodynamics is also given. An
interesting outcome of our analysis is the restriction obtained on
the magnitude of acceleration for these black holes.}

\begin{document}

\section{The Pleba\'{n}ski-Demia\'{n}ski family of black holes}

Pleba\'{n}ski and Demia\'{n}ski \cite{1.22} presented a large class
of solutions of Einstein's equations with a possibly non-zero
cosmological constant $\Lambda $, which includes, apart from other
interesting solutions, the famous Kerr-Newman rotating black hole
and hence the Kerr, the Reissner-Nordstr\"{o}m, and the
Schwarzschild black holes. This family includes, in particular,
solutions for accelerating black holes also. The general form of the
metric thus contains seven free parameters which characterize the
mass $m$, electric and magnetic charges $e$ and $g$ respectively,
Kerr-like rotation parameter $a$ which is equal to angular momentum
per unit mass i.e. $a=J/m$, the NUT (Newman-Unti-Tamburino)
parameter $l$, acceleration of the source $\alpha $ and the
cosmological constant $\Lambda $. We write the general
Pleba\'{n}ski-Demia\'{n}ski metric in the  notation used in Refs.
~\cite{1, 2} as
\begin{eqnarray}\nonumber
ds^{2} &=&\frac{1}{\bar{\Omega}^{2}}\{-\frac{Q}{\rho^{2}}[dt-(a\sin
^{2}\theta +4l\sin ^{2}\frac{\theta
}{2})d\phi]^{2}+\frac{\rho^{2}}{Q}
dr^{2}+\frac{\rho ^{2}}{P}d\theta^{2}   \\
&&+ \frac{P}{\rho ^{2}}\sin ^{2}\theta \lbrack
adt-(r^{2}+(a+l)^{2})d\phi ]^{2}\},  \label{2.4}
\end{eqnarray}
where
\begin{eqnarray}\nonumber
\bar{\Omega}&=&1-\frac{\alpha }{\omega }(l+a\cos \theta )r , \\
\nonumber \rho ^{2}&=&r^{2}+(l+a\cos \theta )^{2} ,\\
\nonumber P&=&1-a_{3}\cos \theta -a_{4}\cos ^{2}\theta ,\\
\nonumber Q&=&(\omega ^{2}k+e^{2}+g^{2})-2mr+\epsilon r^{2}-2\alpha
\frac{n}{\omega} r^{3}-(\alpha ^{2}k+\frac{\Lambda }{3})r^{4} , \\
\nonumber a_{3}&=&2\alpha \frac{a}{\omega }m\ -\ 4\alpha
^{2}\frac{al}{\omega
^{2}} (\omega ^{2}k+e^{2}+g^{2})-4\frac{\Lambda }{3}al , \\
\nonumber a_{4}&=&-\alpha ^{2}\frac{a^{2}}{\omega ^{2}}(\omega
^{2}k+e^{2}+g^{2})-\frac{ \Lambda }{3}a^{2} , \\
\epsilon &=&\frac{\omega ^{2}k}{a^{2}-l^{2}}+4\alpha
\frac{l}{\omega} m-(a^{2}+3l^{2})[\frac{\alpha ^{2}}{\omega
^{2}}(\omega
^{2}k+e^{2}+g^{2})+ \frac{\Lambda }{3}] ,  \label{2.3a} \\
n&=&\frac{\omega ^{2}kl}{a^{2}-l^{2}}-\alpha \frac{a^{2}-l^{2}}{
\omega }m+(a^{2}-l^{2})l[\frac{\alpha ^{2}}{\omega ^{2}}(\omega
^{2}k+e^{2}+g^{2})+\frac{\Lambda }{3}] , \label{2.3b} \\
k&=&[1+2\alpha \frac{l}{\omega }m-3\alpha ^{2}\frac{l^{2}}{\omega
^{2}} (e^{2}+g^{2})-l^{2}\Lambda ](\frac{\omega
^{2}}{a^{2}-l^{2}}+3\alpha ^{2}l^{2})^{-1} . \label{2.3c}
\end{eqnarray}

Here $\epsilon$ and $k$ are arbitrary real parameters, $n$ is the
Pleba\'{n}ski-Demia\'{n}ski parameter and $\omega $ is the twist. If
$\Lambda =0$ or $\Lambda >0$ then for the non-accelerating case the
metric will represent a single black hole, and for the accelerating
case the metric will represent a pair of causally separated black
holes which are accelerating away from each other in opposite
directions. If $\Lambda <0$ then the metric will represent a single
black hole for small acceleration and a pair of black holes if the
acceleration is large.

When $\alpha =0=l=g=\Lambda $ the line element reduces to the
Kerr-Newman solution. Further, we get the Schwarzschild metric if
the electric charge and rotation parameter vanish i.e. $e=0=a$.
Therefore, the line element (\ref{2.4}) is a very convenient metric
representation of the complete class of accelerating, rotating and
charged black holes. The metric is singularity free if $\left\vert
a\right\vert <\left\vert l\right\vert ,$\ and it has a Kerr-like
ring singularity at $\rho =0$ when $\left\vert a\right\vert \geq
\left\vert l\right\vert $.

The rotating and accelerating black holes give rise to two types of
horizons: the rotation horizons (analogous to the Kerr-Newman
horizons) and two acceleration horizons. In this paper we study the
thermodynamical properties of rotating black holes with NUT
parameter and, rotating and accelerating black holes. The form of
ergospheres for these black holes is given. We provide relations for
their Hawking temperature and entropy. The first law of black hole
thermodynamics is also discussed in the context of these spaces. As
a result of our analysis we find an interesting relation which
restricts the amount of acceleration these black holes can have.

\section{Thermodynamics of the non-accelerating black holes}

It can be seen from Eq. (\ref{2.4}) that, when $\alpha =0=\Lambda$,
we have $\omega ^{2}k=a^{2}-l^{2}$ and hence $\epsilon =1$, $ n=l$,
$P=1$, and we get \cite{2}

\begin{eqnarray}\nonumber
ds^{2} =\frac{Q}{\rho ^{2}}[dt-(a\sin ^{2}\theta +4l\sin
^{2}\frac{\theta
}{2})d\phi ]^{2}-\frac{\rho ^{2}}{Q}dr^{2}-\rho ^{2}d\theta ^{2}   \\
-\frac{\sin ^{2}\theta }{ \rho ^{2}}[adt-(r^{2}+(l+a)^{2})d\phi
]^{2} , \label{2.6}
\end{eqnarray}
where
\begin{equation}
\rho ^{2}=r^{2}+(l+a\cos \theta )^{2},
Q=(a^{2}-l^{2}+e^{2}+g^{2})-2mr+r^{2} ,  \label{2.6a}
\end{equation}
which is the Kerr-Newman-NUT solution. We note that $Q=0$ gives the
expression for locations of inner and outer horizons of the black
hole as \cite{2}
\begin{equation}
r_{\pm }=m\pm \sqrt{m^{2}+l^{2}-a^{2}-e^{2}-g^{2}},  \label{2.6b}
\end{equation}
where $m^{2}\geq a^{2}+e^{2}+g^{2}-l^{2}.$

Here we discuss the formation of ergospheres in these black holes.
We know that ergosphere are characterized by \cite{12}
\begin{equation}
g_{tt}=0 .
\end{equation}
So from Eqs. (\ref{2.6}) and (\ref{2.6a})
\begin{equation}
r^{2}-2mr-l^{2}+e^{2}+g^{2}+a^{2}\cos ^{2}\theta =0 .
\end{equation}
Its solution is
\begin{equation}
r_{n}(\theta )=m+\sqrt{m^{2}+l^{2}-e^{2}-g^{2}-a^{2}\cos ^{2}\theta
}  ,
\end{equation}
which gives the ergosphere for the black hole. Now we see its
relation with the outer horizon (\ref{2.6b}). For this we consider
\begin{equation}
0\leq \cos ^{2}\theta \leq 1 ,
\end{equation}
which allows us to write

\begin{equation}
m^{2}+l^{2}-e^{2}-g^{2}-a^{2}\leq m^{2}+l^{2}-e^{2}-g^{2}-a^{2}\cos
^{2}\theta \leq m^{2}+l^{2}-e^{2}-g^{2} ,
\end{equation}
\begin{eqnarray}\nonumber
m+\sqrt{m^{2}+l^{2}-e^{2}-g^{2}-a^{2}}\\ \nonumber \leq m+\sqrt{
m^{2}+l^{2}-e^{2}-g^{2}-a^{2}\cos ^{2}\theta} \\
 \leq m+\sqrt{m^{2}+l^{2}-e^{2}-g^{2}} ,
\end{eqnarray}
or

\begin{equation}
r_{+}\leq r_{n}(\theta )\leq m+\sqrt{m^{2}+l^{2}-e^{2}-g^{2}} ,
\end{equation}
\begin{equation}
r_{+}\leq r_{n}(\theta )\leq r_{a} ,
\end{equation}
where $r_{a}$ is the outer horizon of the corresponding
Reissner-Nordstr\"{o}m black hole with magnetic and NUT charges $g$
and $l$ respectively. The above relation has a beautiful information
to interpret. Since the ergosphere is dependent on $\theta $, so it
will coincide with the outer horizon at $\theta =0$ and stretches
out for other values of $\theta $. However, it cannot stretch beyond
the outer horizon of the corresponding Reissner-Nordstr\"{o}m black
hole. At $\theta =\pi /2$, however, they coincide.

In order to discuss thermodynamics of these black holes with the NUT
parameter, we observe that for a metric of the form
\begin{equation}
ds^{2}=-F(r,\theta )dt^{2}+\frac{1}{G(r,\theta
)}dr^{2}+r^{2}(d\theta ^{2}+\sin ^{2}\theta d\phi ^{2}) ,
\label{3.04}
\end{equation}
the surface gravity, $\kappa$ is given by \cite{7}
\begin{equation}
\kappa =\frac{1}{\sqrt{-h}}\frac{\partial }{\partial x^{a}}\left(
\sqrt{-h}  h^{ab} \frac{\partial r}{\partial x^{b}}\right)  ,
\label{3.05}
\end{equation}
where $h_{ab}$ is the second order diagonal metric made from $t-r$
sector of the metric and $h=\det h_{ab }$. Putting the values of$\
h^{11}$ and $\sqrt{-h}$\ from metric (\ref{2.6}) we get
\begin{equation}
\kappa =\frac{1}{\sqrt{1-\frac{a^{2}\sin ^{2}\theta
}{Q}}}\frac{\partial f(r,\theta )}{\partial r}\ ,  \label{3.5}
\end{equation}
where
\begin{equation}
f(r,\theta )=\frac{Q}{\rho ^{2}}\sqrt{1-\frac{a^{2}\sin ^{2}\theta
}{Q}}  .  \label{3.6}
\end{equation}
Thus Eq. (\ref{3.5}) takes the form
\begin{equation}
\kappa =\frac{1}{\rho ^{2}}\left( 2(r-m)-\frac{2rQ}{\rho
^{2}}+\frac{ a^{2}\sin ^{2}\theta (r-m)}{(Q-a^{2}\sin ^{2}\theta
)}\right)  . \label{3.6c}
\end{equation}
At horizon $Q=0$, $r\rightarrow r_{+}$ and using Eq. (\ref{2.6b}) at
$\theta =0$, this becomes
\begin{equation}
\kappa _{h}=\frac{(r_{+}-m)}{\rho ^{2}}=\frac{(r_{+}-m)}{
[r_{+}^{2}+(l+a)^{2}]} .  \label{3.8}
\end{equation}
or
\begin{equation}
\kappa _{h}=\frac{\sqrt{m^{2}+l^{2}-a^{2}-e^{2}-g^{2}}}{
2m^{2}+2l^{2}+2al-e^{2}-g^{2}+2m\sqrt{m^{2}+l^{2}-a^{2}-e^{2}-g^{2}}}
, \label{3.9}
\end{equation}
It is important to note that if we put the NUT parameter $l=0$\ and
the magnetic charge $g=0$, then Eq. (\ref{3.9}) reduces to the
surface gravity for the Kerr-Newman black hole. Further if \ $e=0=a$
then the relation for the Schwarzschild black hole is obtained.

The relation (\ref{3.9}) can also be written in terms of the inner
and outer horizons by noting that
\begin{equation}
(r_{+}-m)-(r_{-}-m)=2(r_{+}-m) .
\end{equation}
Thus Eq. (\ref{3.8}) becomes
\begin{equation}
\kappa _{h}=\frac{r_{+}-r_{-}}{2[r_{+}^{2}+(l+a)^{2}]} .
\label{3.9b}
\end{equation}
We can also find the surface gravity by using the angular velocity
\cite{6}
\begin{equation}
\Omega =\frac{d\phi }{dt}=-\frac{g_{t\phi }}{g_{\phi \phi }} .
\label{3.9c}
\end{equation}
So we have
\begin{equation}
\Omega _{h}=\frac{a}{r_{+}^{2}+(l+a)^{2}} .
\end{equation}
or
\begin{equation}
\Omega _{h}=\frac{a}{2m^{2}+2l^{2}+2al-e^{2}-g^{2}+2m\sqrt{
m^{2}+l^{2}-a^{2}-e^{2}-g^{2}}} .  \label{3.12}
\end{equation}
This is the angular velocity for the non-accelerating black hole.
The formula for surface gravity in terms of angular velocity is
\cite{10}
\begin{equation}
\kappa _{h}=\frac{1}{2a}\Omega _{h}\left. \frac{dQ}{dr}\right\vert
_{_{r=r_{+}}} .  \label{3.12a}
\end{equation}
Using Eq. (\ref{3.12}) and $dQ/dr=2(r_{+}-m),$ we get the same
result as Eq. (\ref {3.9}).


As we know that the temperature of a black hole is given by
\cite{12}
\begin{equation}
T=\frac{\kappa _{h}}{2\pi } .  \label{3.023a}
\end{equation}
Putting the value of $\kappa _{h}$ from Eq. (\ref{3.9}) we get
\begin{equation}
T=\frac{1}{2\pi }\left[ \frac{\sqrt{m^{2}+l^{2}-a^{2}-e^{2}-g^{2}}}{
2m^{2}+2l^{2}+2al-e^{2}-g^{2}+2m\sqrt{m^{2}+l^{2}-a^{2}-e^{2}-g^{2}}}\right]
 ,  \label{3.10}
\end{equation}
which is the temperature for the non-accelerating black holes. The
temperature for the Kerr-Newman black hole can directly be deduced
by putting $l=0=g$. Let us see the behavior of the temperature with
the mass graphically.


\FIGURE{\epsfig{file=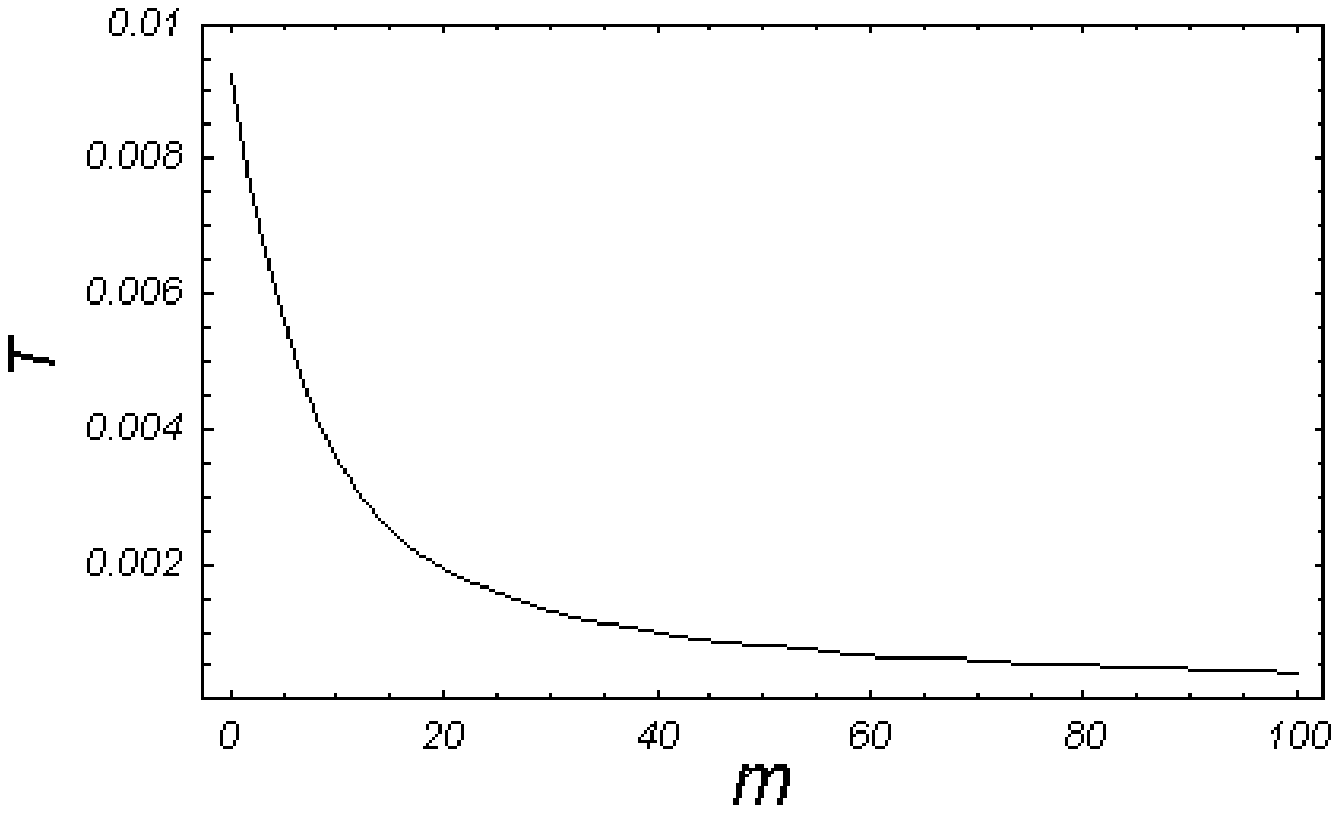,width=10cm}
        \caption[Example of figure]{Temperature versus mass for the
Kerr-Newman-NUT black hole when $l=5, \alpha = e = g = 2$.}
 \label{fig1}}

\FIGURE{\epsfig{file=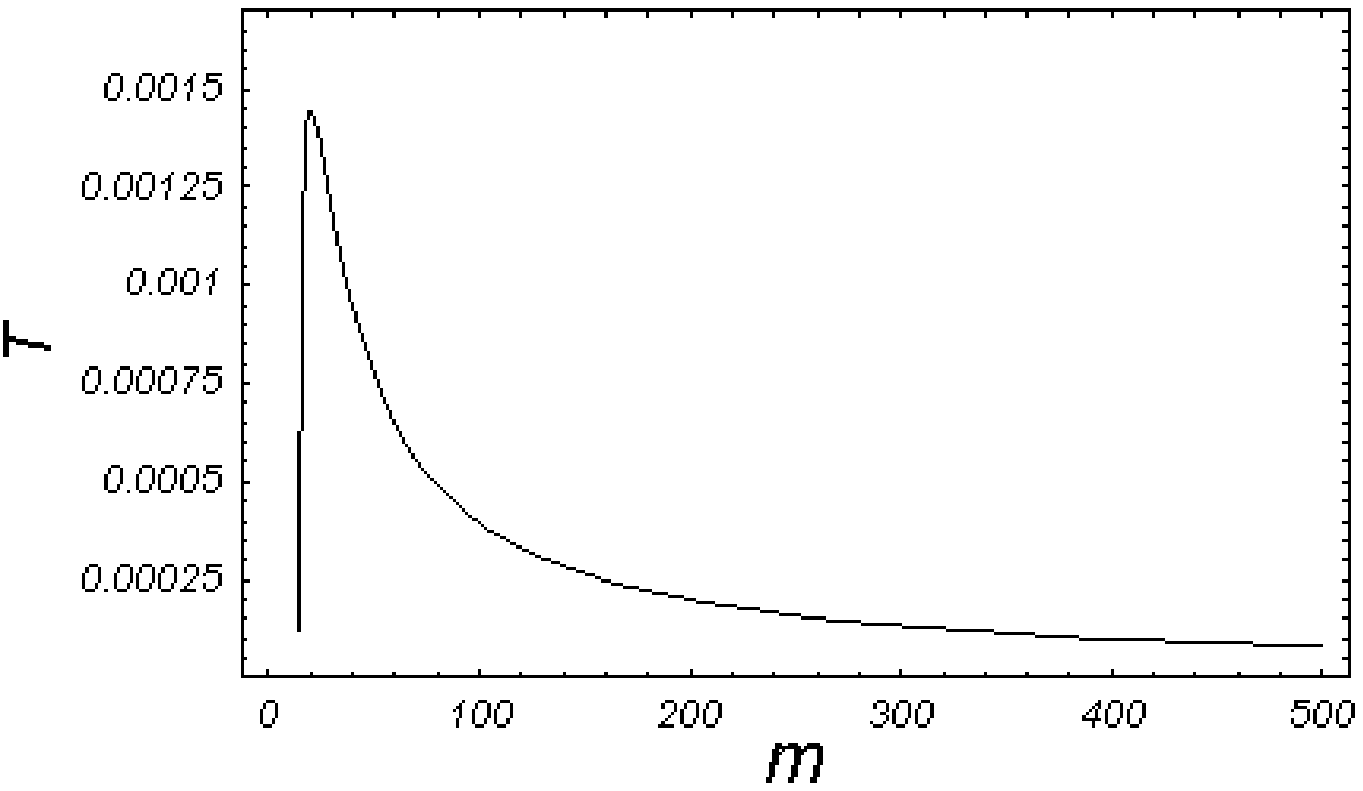,width=10cm}
        \caption[Example of figure]{Temperature versus mass for the
Kerr-Newman-NUT black hole when $l=9, \alpha = e = g = 10$.}
 \label{fig1}}

We see from Figures 1 and 2 that if $l^{2}>a^{2}+e^{2}+g^{2}$ then
the temperature decreases with the increasing mass but will never be
zero. If $ l^{2}<a^{2}+e^{2}+g^{2}$, the temperature will first
increase when $m>\sqrt{ a^{2}+e^{2}+g^{2}-l^{2}}$ to the value of
$m$ where $dT/dm=0$ and then it will decrease with the increasing
mass. In this case temperature shows the same behavior as that of
the Kerr-Newman black hole. We further note that for a fixed value
of the NUT parameter, the maximum value of the temperature is
inversely proportional to the collective magnitude of the rotation
parameter and the electric and magnetic charges.

The horizon area of a rotating black hole is defined as \cite{12}
\begin{equation}
A=\frac{4\pi a}{_{\Omega _{h}}} .  \label{3.10a}
\end{equation}
Substituting from Eq. (\ref{3.12}), this becomes
\begin{equation}
A=4\pi \left[ 2(m^{2}+l^{2}+al+m\sqrt{m^{2}+l^{2}-a^{2}-e^{2}-g^{2}}
)-e^{2}-g^{2}\right]  .  \label{3.13}
\end{equation}


The entropy of a black hole \cite{12}
\begin{equation}
S=\frac{A}{4} ,  \label{3.13a}
\end{equation}
takes the form
\begin{equation}
S=\pi \left[ 2(m^{2}+l^{2}+al+m\sqrt{m^{2}+l^{2}-a^{2}-e^{2}-g^{2}}
)-e^{2}-g^{2}\right]  .  \label{3.14}
\end{equation}


 \FIGURE{\epsfig{file=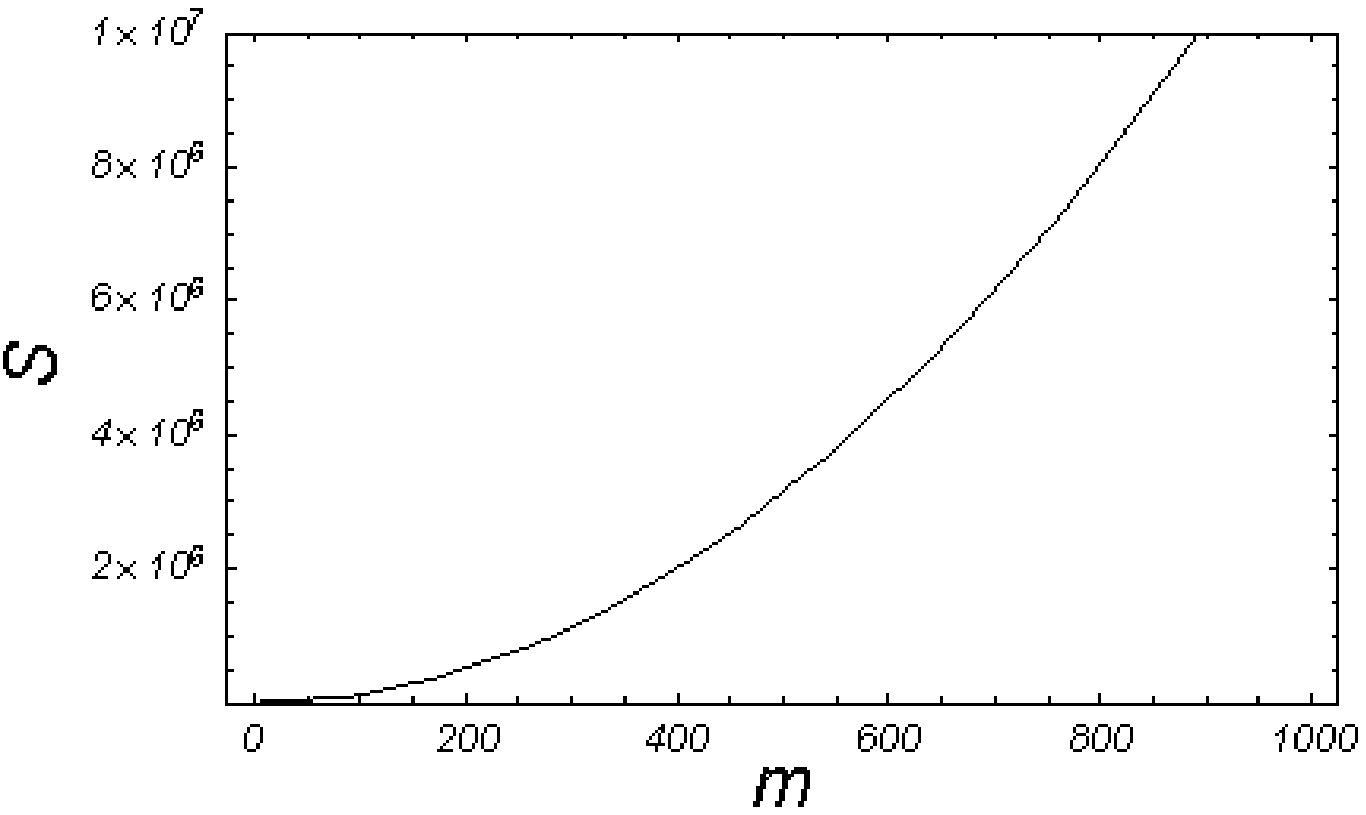,width=10cm}
        \caption[Example of figure]{Entropy versus mass for the
Kerr-Newman-NUT black hole when $l=2, \alpha = e = g = 3$.}
 \label{fig1}}

From Figure 3 we see that entropy is an increasing function of mass
in accordance with the second law of black hole thermodynamics. The
standard formula for the entropy of the Kerr-Newman black hole can
be recovered from this.


Now, the first law of thermodynamics in the form of the law of
conservation of mass is given by \cite{12, 14}
\begin{equation}
dm=\frac{\kappa _{h}}{8\pi }dA+\Omega _{h}dJ+\Phi _{h}de ,
\label{3.15}
\end{equation}
where $\Phi _{h}$\ is the electrostatic potential of the black hole
\cite{12}
\begin{equation}
\Phi _{h}=\frac{4\pi er_{+}}{A} .  \label{3.15a}
\end{equation}
Putting the value of $r_{+}$ from Eq. (\ref{2.6b}) and of horizon
area from Eq. (\ref{3.13}), it becomes
\begin{equation}
\Phi _{h}=\frac{e[m+\sqrt{m^{2}+\eta
}]}{2(m^{2}+l^{2}+al+m\sqrt{m^{2}+\eta } )-e^{2}-g^{2}} ,
\end{equation}
where\ $\eta =l^{2}-a^{2}-e^{2}-g^{2}$ .

Substituting the values of $\kappa _{h}$ from Eq. (\ref{3.9}),
$\Omega _{h}$ from Eq. (\ref{3.12}) and $\Phi _{h}$ from above in
Eq. (\ref{3.15}), we get
\begin{equation}
dm=\frac{1}{\mu }\left[ \frac{\sqrt{m^{2}+\eta }}{8\pi
}dA+adJ+e(m+\sqrt{ m^{2}+\eta })de\right]  ,  \label{3.16}
\end{equation}
where$\ \mu =2m^{2}+2l^{2}+2al-e^{2}-g^{2}+2m\sqrt{m^{2}+\eta }$ .

This is the first law of thermodynamics for the non-accelerating
black holes.

\section{Thermodynamics of accelerating and rotating black holes}

As mentioned earlier the Pleba\'{n}ski-Demia\'{n}ski metric includes
the rotating and accelerating charged black holes. Now we consider
another form of the metric which is free of NUT-like behavior i.e.
we take $l=0$. If we put $l=0,\ k=1,$ in Eq. (\ref{2.4}) then\
$\omega =a, a_{3}=2\alpha m,\ a_{4}=-\alpha
^{2}(a^{2}+e^{2}+g^{2})-\frac{\Lambda }{3}a^{2}$ and substituting
for $\epsilon $ and $n$, the\ line element (\ref{2.4}) with $
\Lambda \neq 0$ will take the form \cite{5}

\begin{equation}
ds^{2}=\frac{1}{\bar{\Omega}^{2}} \{-\frac{\bar{Q}}{\rho
^{2}}[dt-a\sin ^{2}\theta d\phi ]^{2}+\frac{P}{\rho ^{2}}\sin
^{2}\theta \lbrack adt-(r^{2}+a^{2})d\phi ]^{2}+\frac{\rho
^{2}}{\bar{Q}}dr^{2}+\frac{\rho ^{2} }{P}d\theta ^{2} \} ,
\label{2.7}
\end{equation}
where

\begin{equation}
\bar{\Omega}=1-\alpha r\cos \theta  , \rho ^{2}=r^{2}+a^{2}\cos
^{2}\theta  ,  \label{2.7a}
\end{equation}
\begin{equation}
P=1-2\alpha m\cos \theta +\{\alpha
^{2}(a^{2}+e^{2}+g^{2})+\frac{\Lambda a^{2}}{3}\}\cos ^{2}\theta
 ,
\end{equation}
\begin{equation}
\bar{Q}=\{(a^{2}+e^{2}+g^{2})-2mr+r^{2}\}(1-\alpha
^{2}r^{2})-\frac{\Lambda }{3}(r^{2}+a^{2})r^{2} .  \label{a1}
\end{equation}
The above metric contains six arbitrary parameters $m,\ e,\ g,\
\alpha ,\ a$ and $\Lambda $ which can be varied independently. We
will not consider the cosmological constant in our discussion so we
will put this equal to zero.

Here $\rho ^{2}=0$ indicates the presence of a Kerr-like ring
singularity at $r=0$ and $\theta =\pi /2$.  In this case $\bar{Q}=0$
gives the expression for the locations of inner and outer horizons,
which are identical to those of the non-accelerating Kerr-Newman
black hole \cite{5}
\begin{equation}
r_{\pm }=m\pm \sqrt{m^{2}-a^{2}-e^{2}-g^{2}} ,  \label{a1a}
\end{equation}
where $a^{2}+e^{2}+g^{2}\leq m^{2}$. However, in addition to these,
there are acceleration horizons also at $r=1/\alpha $ and
$r=1/\alpha \cos \theta$, which come from putting $\bar{Q}=0$ and
$\bar{\Omega}=0$ respectively, and are coincident with each other at
$\theta =0.$


We first find the surface gravity of the accelerating and rotating
black holes using Eq. (\ref{3.05}). For $\Lambda =0$, from Eq.
(\ref{a1}) we have
\begin{equation}
\bar{Q}=(r^{2}-2mr+a^{2}+e^{2}+g^{2})(1-\alpha ^{2}r^{2}) .
\label{3.20a}
\end{equation}
Thus Eq. (\ref{3.05}) becomes
\begin{equation}
\kappa =\frac{\bar{\Omega}^{2}}{\rho ^{2}}\left[
\frac{1}{2}\frac{d\bar{Q}}{ dr}\{1+\frac{\bar{Q}}{\bar{Q}-Pa^{2}\sin
^{2}\theta }\}-\frac{1}{\rho ^{2}}2r \bar{Q}\right]  . \label{3.22}
\end{equation}
At horizon, $\bar{Q}=0$, therefore, Eq. (\ref{3.22}) becomes
\begin{equation}
\kappa _{h}=\frac{\bar{\Omega}^{2}}{2\rho ^{2}}\left[
\frac{d\bar{Q}}{dr} \right]  .
\end{equation}
As $\ \bar{\Omega}\neq 0$ at $r=r_{+}$ we get
\begin{equation}
\kappa _{h}=\frac{\bar{\Omega}^{2}}{2\rho ^{2}}2[(r-m)(1-\alpha
^{2}r^{2})-\alpha ^{2}r(r^{2}-2mr+a^{2}+e^{2}+g^{2})] .
\end{equation}
Since at outer horizon $r^{2}-2mr+a^{2}+e^{2}+g^{2}=0$, therefore,
putting the value of $\rho ^{2}$ and $\bar{ \Omega}$ from Eq.
(\ref{2.7a}) at $\theta =0$, we get
\begin{equation}
\kappa _{h}=\frac{(r_{+}-m)}{(r_{+}^{2}+a^{2})}(1-\alpha
r_{+})^{3}(1+\alpha r_{+}) .  \label{3.26}
\end{equation}
Using Eq. (\ref{a1a}) this takes the form
\begin{equation}
\kappa _{h}=\frac{[1-\alpha
(m+\sqrt{m^{2}-a^{2}-e^{2}-g^{2}})]^{3}[1+\alpha
(m+\sqrt{m^{2}-a^{2}-e^{2}-g^{2}})]}{(\sqrt{m^{2}-a^{2}-e^{2}-g^{2}}
)^{-1}[2m^{2}-e^{2}-g^{2}+2m\sqrt{m^{2}-a^{2}-e^{2}-g^{2}}]} ,
\label{3.27}
\end{equation}
which is the surface gravity for the accelerating and rotating black
holes at the outer horizon. Note that from Eq. (\ref{3.26}) the
surface gravity will vanish at the acceleration horizon,
$r=1/\alpha$ (and also for the other acceleration horizon at $\theta
=0$). In the above equation if $\alpha =0$, it reduces to the
relation of surface gravity for the non-accelerating case (Eq.
(\ref{3.9})) at $l=0$. Further if $g=0$ the surface gravity for the
Kerr-Newman black hole is obtained.

The relation (\ref{3.27}) can also be written in terms of the inner
and outer horizons. For this we use the relation
$r_{+}-r_{-}=(r_{+}-m)-(r_{-}-m)=2(r_{+}-m)$ in Eq. (\ref{3.26}) to
get
\begin{equation}
\kappa _{h}=\frac{r_{+}-r_{-}}{2(r_{+}^{2}+a^{2})}(1-\alpha
r_{+})^{3}(1+\alpha r_{+}) .  \label{3.27a}
\end{equation}

Now, in this case the angular velocity from Eq. (\ref{3.9c}) becomes
\begin{equation}
\Omega =\frac{a[\bar{Q}-P(r^{2}+a^{2})]}{\bar{Q}a^{2}\sin ^{2}\theta
-P(r^{2}+a^{2})^{2}} .
\end{equation}
At horizon, $\bar{Q}=0$, thus the angular velocity for the
accelerating and rotating black holes can be written as
\begin{equation}
\Omega
_{h}=\frac{a}{2m^{2}-e^{2}-g^{2}+2m\sqrt{m^{2}-a^{2}-e^{2}-g^{2}}}
 .  \label{3.29}
\end{equation}
Since angular velocity is calculated at the outer horizon which is
not dependent on acceleration, therefore, this is the same as that
of the non-accelerating case. Using this and Eq. (\ref{3.12a}) we
get the same formula for the surface gravity as given in Eq.
(\ref{3.27}), and by $T=\kappa _{h}/2\pi$ we get the temperature for
the accelerating and rotating black holes. We see that the factor
$[1-\alpha (m+\sqrt{m^{2}-a^{2}-e^{2}-g^{2}})]$ can make the
temperature negative for large values of $\alpha$, so we have to
allow only small magnitudes of acceleration. Note that the
temperature vanishes at
\begin{equation}
1-\alpha (m+\sqrt{m^{2}-a^{2}-e^{2}-g^{2}})=0 ,
\end{equation}
or
\begin{equation}
\alpha =\frac{1}{m+\sqrt{m^{2}-a^{2}-e^{2}-g^{2}}} ,
\end{equation}
and in terms of mass we have
\begin{equation}
m=\frac{1+\alpha ^{2}(a^{2}+e^{2}+g^{2})}{2\alpha } .
\end{equation}
Thus to avoid the cases of extremal black holes and the violation of
the third law of black hole thermodynamics, we need to restrict the
values of $\alpha$. We find that the permitted range is
\begin{equation}
\alpha <\frac{1}{m+\sqrt{m^{2}-a^{2}-e^{2}-g^{2}}} .
\end{equation}

\FIGURE{\epsfig{file=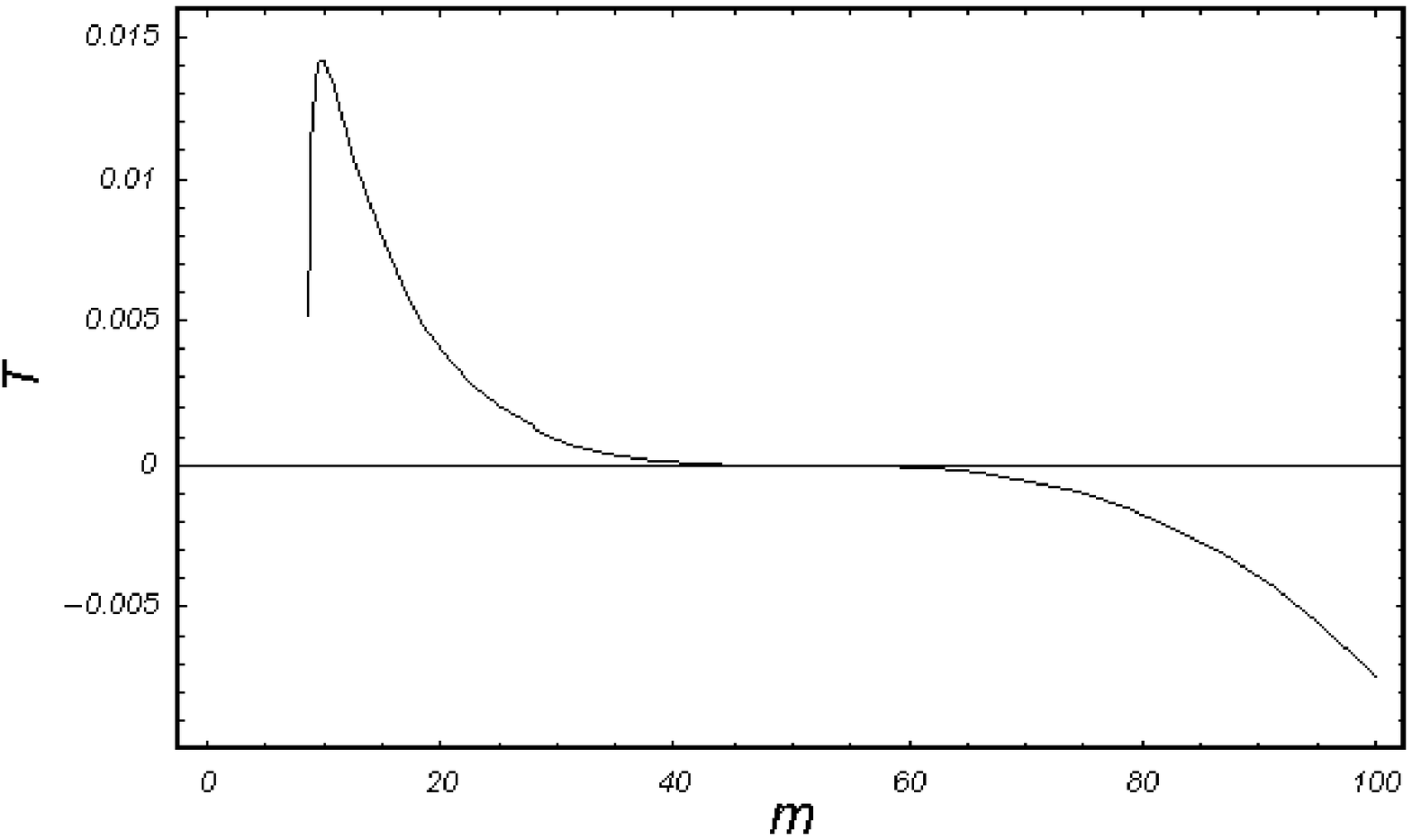,width=10cm}
        \caption[Example of figure]{Temperature versus mass for the accelerating
and rotating black hole when $\alpha = 0.01, a = e = g = 5$.}
 \label{fig1}}


From Figure 4 we see that, for small values of $\alpha$, the
temperature of these black holes shows the same behavior as that of
the non-accelerating black holes. When the temperature becomes less
than zero the black hole obviously ceases to exist. One can see that
smaller is the acceleration greater is the maximum value of the
temperature. Similarly, lesser the electric and magnetic charges and
the rotation parameter greater is this maximum value.

In order to write the first law of thermodynamics, we note that the
horizon area of the black hole using angular velocity from Eq.
(\ref{3.29}) is given by
\begin{equation}
A=4\pi \lbrack 2m^{2}-e^{2}-g^{2}+2m\sqrt{m^{2}-a^{2}-e^{2}-g^{2}}]
. \label{3.30}
\end{equation}

Using the values of $r_{+}$ and $A$\ from Eqs. (\ref{a1a}) and
(\ref{3.30}) the electrostatic potential from Eq. (\ref{3.15a}) can
be written as
\begin{equation}
\Phi
_{h}=\frac{e[m+\sqrt{m^{2}-\bar{\eta}}]}{[2m^{2}-e^{2}-g^{2}+2m\sqrt{
m^{2}-\bar{\eta}}]} ,  \label{3.32}
\end{equation}
where $\bar{\eta}=a^{2}+e^{2}+g^{2}.$

Now putting the values of$\ \kappa _{h}$, $\Omega _{h}$ and $\Phi
_{h}$ from Eqs. (\ref{3.27}), (\ref{3.29}) and (\ref{3.32}) in
(\ref{3.15}), we get the first law for the accelerating and rotating
black holes as
\begin{eqnarray}\nonumber
dm &=&\frac{1}{\bar{\mu}}[\frac{\sqrt{m^{2}-\bar{\eta}}\ [1-\alpha
(m+\sqrt{ m^{2}-\bar{\eta}} )]^{3}[1+\alpha
(m+\sqrt{m^{2}-\bar{\eta}} )]
}{8\pi }dA  \\
&&+adJ+e(m+\sqrt{ m^{2}-\bar{\eta}})de] ,  \label{3.33}
\end{eqnarray}
where $\bar{\mu}=2m^{2}-e^{2}-g^{2}+2m\sqrt{m^{2}-\bar{\eta}}$ .

\section{Conclusion}

We have seen that the ergosphere for the Kerr-Newman-NUT black hole
touches its outer horizon and stretches out upto a limit which
corresponds to the horizon of the Reissner-Nordstr\"{o}m black hole.
The temperature for the Kerr-Newman-NUT black hole behaves like that
of the Kerr-Newman black hole if the NUT parameter has less
magnitude than that of the sum of the electric and magnetic charges
along with the rotation parameter and shows a monotonically
decreasing behavior when the NUT parameter has a magnitude more than
the other three mentioned parameters. The maximum value of
temperature is inversely proportional to the collective magnitude of
the rotation parameter and the electric and magnetic charges.

For the accelerating and rotating case, we find that the large
values of acceleration do not represent black holes. Only small
values of acceleration are physically acceptable.

The entropy for both the accelerating and rotating, and the
Kerr-Newman-NUT black holes justify the second law of
thermodynamics. All the results and the thermodynamical quantities
reduce to those of the Kerr-Newman black holes when $l=0=g=\alpha$.
Further, they reduce to the Schwarzschild black holes in appropriate
limits. It is worth mentioning here that the surface gravity and
hence the temperature for accelerating and rotating black holes
vanish on the acceleration horizon and violate the third law of
black hole thermodynamics.

\acknowledgments

Useful discussion with Jiri Podolsk\'{y} is gratefully acknowledged.

\end{document}